\documentclass[structabstract]{aa}
\usepackage{graphicx}
\usepackage{txfonts}
\usepackage{natbib}
\begin{document}
   \title{Velocity vectors of a quiescent prominence observed by \emph{\emph{Hinode}}/SOT and the MSDP (Meudon)}

   \author{B. Schmieder
          \inst{1}
      R. Chandra
      \inst{1}
      A. Berlicki
      \inst{2}
       \and
      P. Mein
     \inst{1}
          }

\institute{Observatoire de Paris, LESIA, UMR8109 (CNRS), F-92195 Meudon Principal Cedex, France\\
              \email{brigitte.schmieder@obspm.fr}
         \and
             Astronomical Institute, Academy of Sciences, Ond\v{r}ejov, Czech Republic\\
 Astronomical institute, University of Wroc{\l}aw, Poland
             }

   \date{Received -------------; accepted -------------}


  \abstract
{The dynamics of prominence fine structures  is a challenge to
understand the formation of cool plasma prominence embedded in the
hot corona.} {Recent observations from  the high resolution \emph{Hinode}/SOT telescope allow us to compute  velocities perpendicularly to the
line-of-sight or transverse velocities. Combining simultaneous
observations obtained in H$\alpha$ with \emph{Hinode}/SOT and the MSDP
spectrograph operating in the Meudon solar tower we derive the
velocity vectors of a quiescent prominence.} {The velocities
perpendicular to the line-of-sight  are measured by time slice
technique, the Dopplershifts by the bisector method.} {The
Dopplershifts of bright threads derived from the MSDP reach 15 km
s$^{-1}$ at the edges of the prominence and are between $\pm$ 5 km
s$^{-1}$ in the center of the prominence. Even though they are minimum
values due to seeing effect, they are
of the same order as the transverse velocities.} {These measurements
are  very important because they suggest that the vertical
structures shown in SOT may not be real vertical magnetic
structures in the sky plane.
The vertical structures could be a pile up of dips in more
or less horizontal magnetic field lines in a 3D perspective,
as it was proposed by many MHD modelers. In our
analysis we also calibrate the \emph{Hinode} H$\alpha$ data using MSDP
observations obtained simultaneously.}

\titlerunning{Velocity vectors of a quiescent prominence}
\authorrunning {B. Schmieder et al.}
   \maketitle
%

\section{Introduction}

The existence of cool structures in so called prominences and filaments
during a few solar rotations embedded in the hot corona has been a
mystery since the beginning of their spectrographic observations
\citep{Azambuja48}. Many reviews concerned the study of quiescent
prominences \citep{Schmieder89,Tandberg94,Labrosse09,Mackay09}.
Since that time period, it is a challenge to derive what is
the best mechanism to succeed to maintain cool plasma in the corona.
A very popular idea is that the plasma is frozen in magnetic field
lines and  stays cool due to the  low transverse thermal conduction
\citep{Demoulin89}. Many magnetic and static models have been
developed on this idea \citep{Kuperus74,Kippenhahn57,Aulanier98,Dudik08}.
But a big question remains: how to get cool plasma
inside the field lines into the corona and how to keep it there?  It is
recognized that this material should come from the chromosphere by
levitation or by injection \citep{Saito73}.
Sufficient mass should be extracted from the chromosphere by
magnetic forces which inject or lift the plasma or by pressure
forces which evaporate the plasma and then cool it to prominence
temperatures. Many models have been developed in this sense, i.e.
thermal non-equilibrium models \citep{Mariska85,Karpen03,Karpen05}.
Levitation models are
proposed through possible magnetic reconnection
\citep{Ballegooijen89}. Injection models could be due to injection
through the reconnection of magnetic field during canceling flux.
These models indicate that the plasma in prominences should have a
large dynamics and static models  may be obsolete. Recent
observations at the Swedish Solar Telescope (SST) show highly dynamic
plasma in filament threads \citep{Lin03,Lin05}. It
was also a first attempt to compute the velocity vectors of the
filament threads. They conclude that the threads inclination from
the horizontal were around 16 degrees with a net flow in both
directions of 8 km s$^{-1}$.  Fine counterstreaming flow is often observed
along horizontal threads or in the barbs \citep{Zirker98,Schmieder91,Schmieder08}.
\emph{Hinode}/SOT movies (available with the electronic version of the paper) reveal  
strong dynamics in the prominence fine structures. The spicules close to the barbs could
allow to inject plasma inside the fine threads. Is that sufficient
to feed continuously the main core of the prominence?  A mass budget
should be done. Using \emph{Hinode} observations  at the limb
\citet{Berger08} and \citet{Chae08} tried to answer  these
questions. They reported different velocity measurements. \citet{Berger08}
found upflows of  dark bubbles around 20 km s$^{-1}$ and down
flows of bright knots less than 10 km s$^{-1}$ looking at Ca II H images in
the line center. \citet{Chae08} analysed an hedgerow prominence
and found horizontal displacements before observing downflows of
bright knots suggesting the existence of  vortex motions.

\begin{figure}
   \centering
\hspace*{-1.5cm}
   \includegraphics[width=0.60\textwidth,clip=]{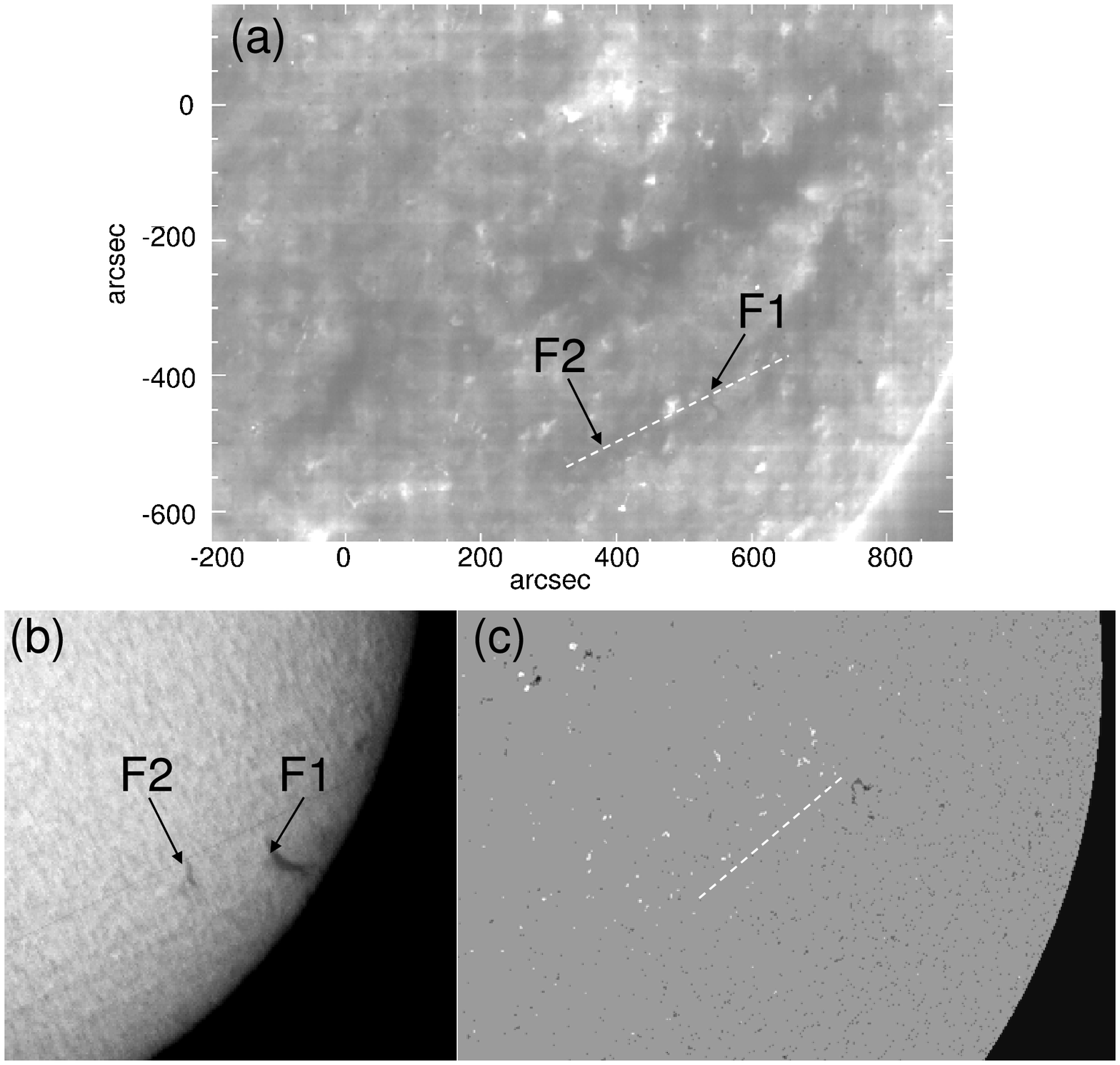}
\caption{(a)  EIT 304 \AA\ image observed on April 20, 2007 at 01:00 UT,
the field of view is 1100$\times$750 arcsec, (b) Filament observed in
H$\alpha$ at Meudon on April 22, 2007 at 15:28 UT, (c) MDI
longitudinal magnetic field on April 21 at 08:00 UT. The magnetic
inversion line is represented by the dashed line in the middle of
the EUV filament channel and in MDI image. The letters F1 and F2
indicate approximately the location of the two filament fragments visible in
H$\alpha$. }\label{meudon}
\end{figure}

Prominence dynamics at the limb look quite different from filament
dynamics on the disk. The integration  along the line of sight
complicates the interpretation. \citet{Mein91} show that more than
15 threads may be integrated along the line of sight and the
resulting velocities should depend on two different Gaussian
distributions. At the edges of prominences the velocity values are
higher because fewer threads are integrated and velocity cells are
larger than intensity knots revealing that {\bf bunches} of threads
may move with the same velocity plasma (Dopplershifts). Similar
results have been  found from completely different approaches.
Therefore to reproduce Lyman line profiles observed by SOHO/SUMER,
\citet{Gunar07} introduce 10 threads perpendicularly to the
line-of-sight with a random velocity distribution in a 2D non LTE
radiative transfer code.

We proposed in this study to study both the velocity perpendicular
to the line of sight using one hour of observations of \emph{Hinode}/SOT in
H$\alpha$ combined with Dopplershifts observed  also in H$\alpha$
with the Multichannel Subtractive Double Pass spectrograph (MSDP)
operating in the Meudon solar tower. These observations were
obtained simultaneously during a coordinated observing program
(JOP178). It is the first time that using \emph{Hinode} data such fine
structures are resolved in prominences and that oscillations and
transverse velocities can be derived \citep{Okamoto07,Berger08,Chae08}.
\emph{Hinode} has a much better
spatial resolution than MSDP by a factor of 5, but the MSDP
observations are very useful to calculate the Dopplershifts and to
calibrate the intensity of  \emph{Hinode}/SOT observations.


\section{Observations}

\begin{figure}
   \centering
   \includegraphics[width=0.45\textwidth,clip=]{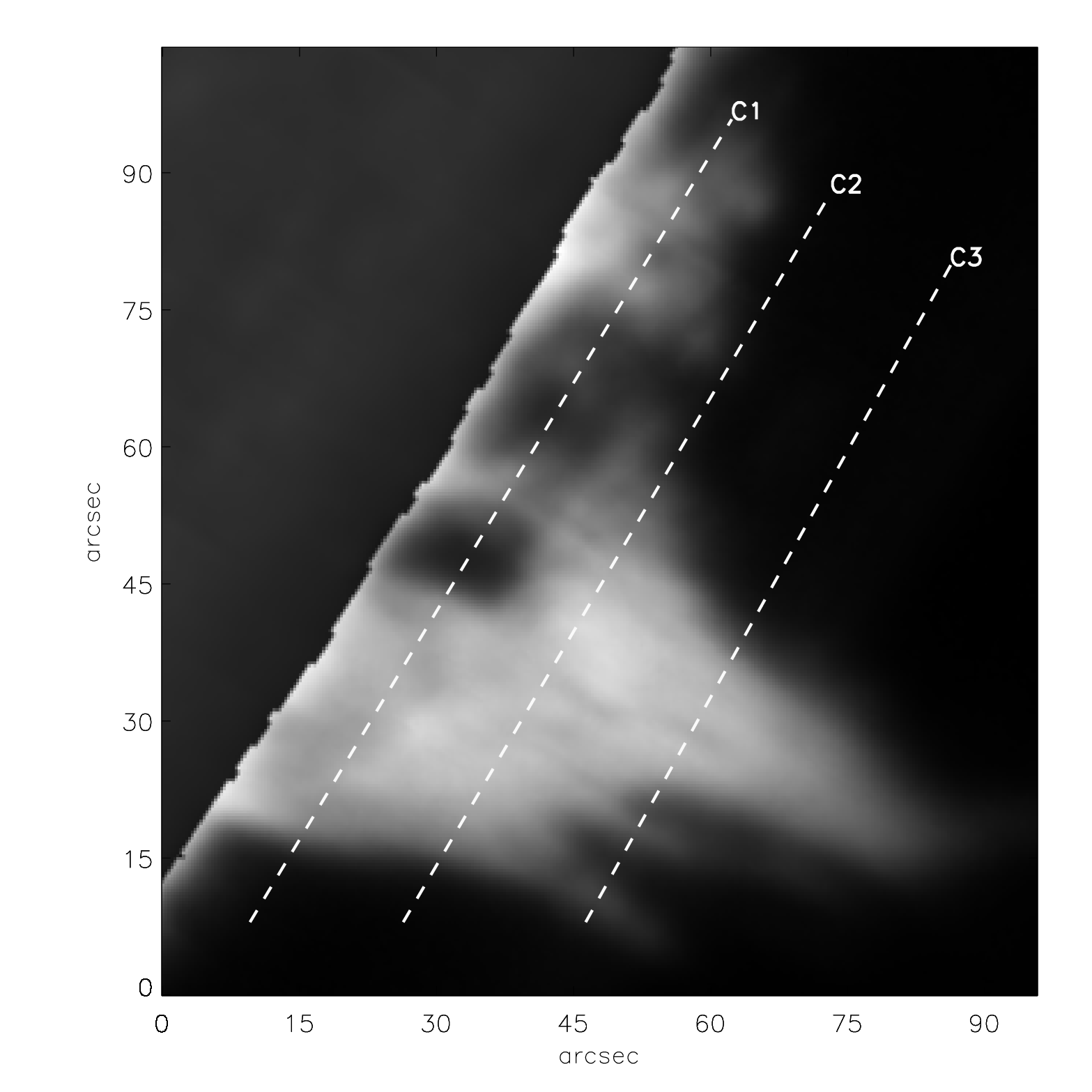}
   \includegraphics[width=0.45\textwidth,clip=]{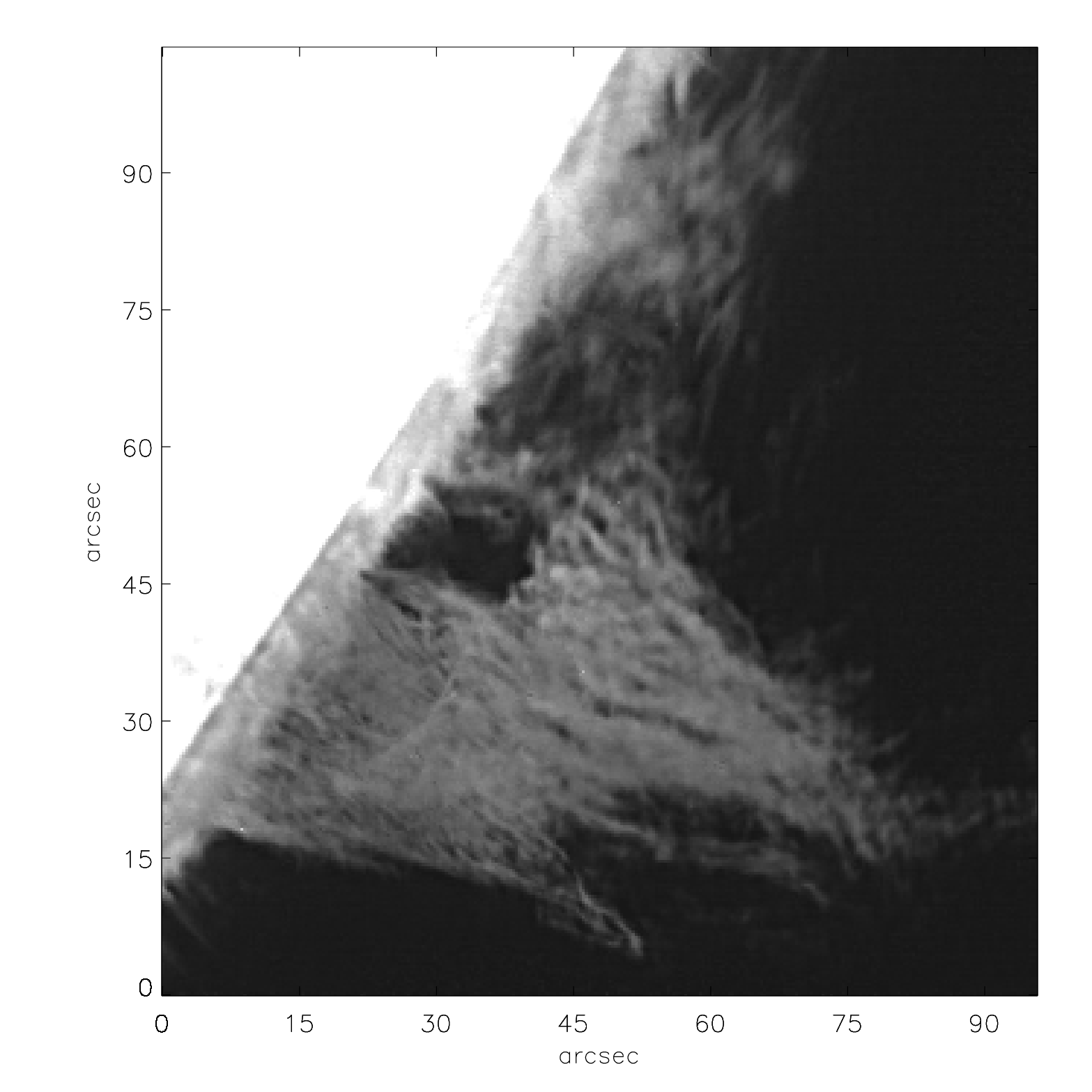}
\caption{Top panel: Observation of the prominence in H$\alpha$ line
center by the  MSDP spectrograph at 13:19:56 UT. Bottom panel:
H$\alpha$  image of prominence with \emph{Hinode}/SOT at 13:19:50 UT on 25
April 2007. The 3 cuts used for the calibration are indicated in the top panel.}\label{msdp_sot}
\end{figure}

The observations presented here were obtained during a coordinated
campaign of prominence observations involving \emph{Hinode}, SOHO
and TRACE missions, as well as several ground-based observatories.
These observations were performed in the JOP 178 framework between
April 23--29 2007 during the first SUMER-\emph{Hinode} observing
campaign. JOP 178 has run successfully many times in the past (see
http://bass2000.obs-mip.fr/jop178/index.html). JOP 178 is dedicated
to the study of prominences and filaments, investigating various
aspects such as their three-dimensional structure and their magnetic
environment from the photosphere to the corona. The observations of
the cavity of the prominence principally by \emph{Hinode}/XRT and
TRACE have been described in details by \citet{torok09} and
\citet{Heinzel08}. The prominence is difficult to observe on the
disk. Two fragments, F1, F2 have been observed on April 21 and 22
2007 in H$\alpha$ survey spectroheliograms of  Meudon (Fig.
\ref{meudon}). On April 21 they are located at S 33, W 40-50 degree
for F1 and S 35 W 35 degree for F2. F1 with a `Y' shape has one
branch  aligned along a parallel and the other is inclined. It
crosses the limb on April 25. The angle P is negative, the leading
part of F1 was in the south compared with the following part F1 as
crossing the limb. These two fragments  are the denser parts of a
filament and look to be the feet, the main body is less dense and
not visible on the survey observations as it is commonly described
\citep{Malherbe89}.  EIT observes a large dark filament channel on
April 21 between   an area of positive magnetic flux in the North
and negative polarities in the South. The longitudinal magnetic flux
is  less than $\pm$ 10 G. It is difficult to derive the polarity
inversion line even when the filament is in the middle of the west
quadrant on April 20 (Fig. \ref{meudon}). This prominence is a very
quiescent prominence.

\begin{figure}
   \centering
\includegraphics[width=0.45\textwidth,clip=]{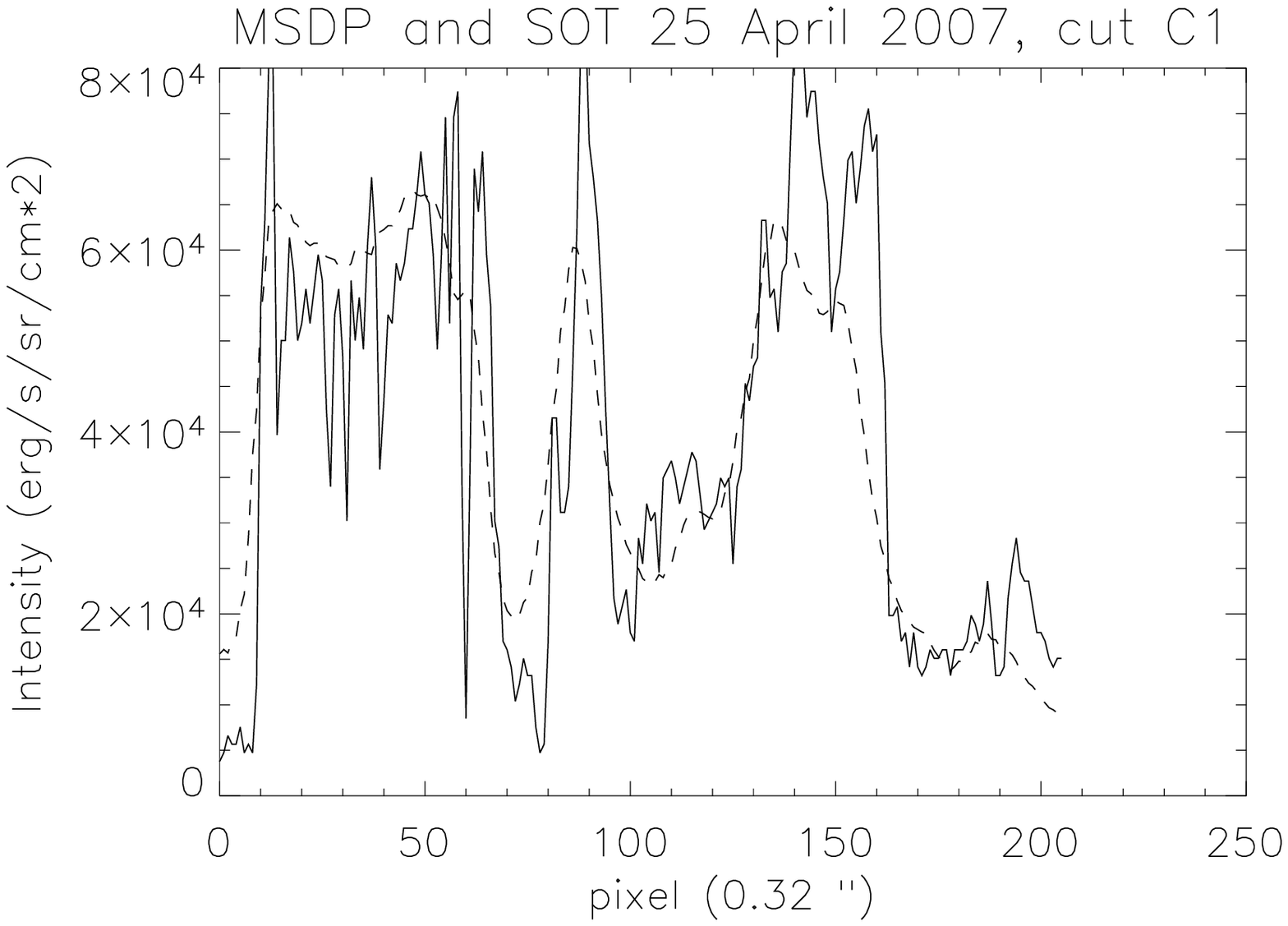}
\includegraphics[width=0.45\textwidth,clip=]{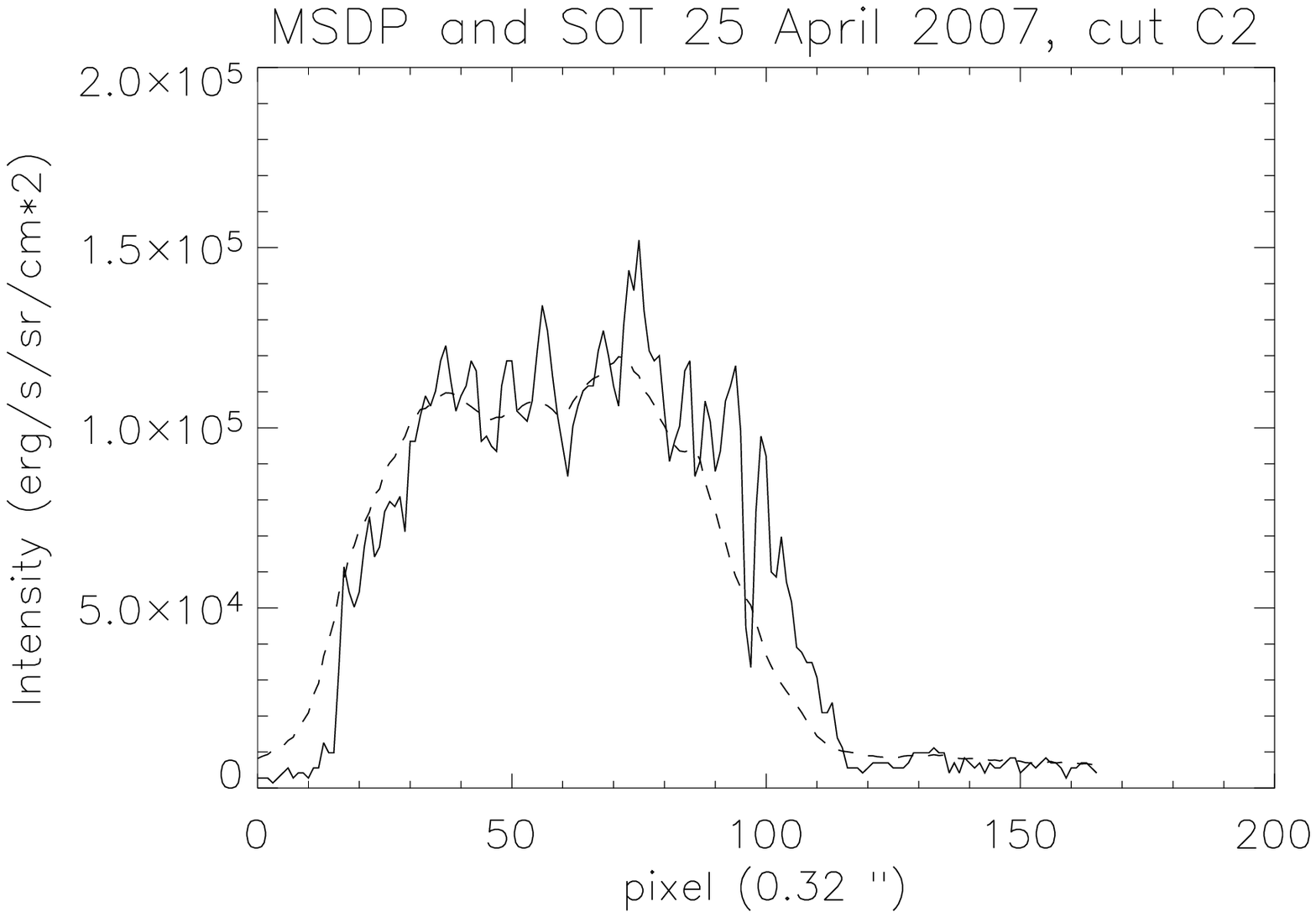}
   \includegraphics[width=0.45\textwidth,clip=]{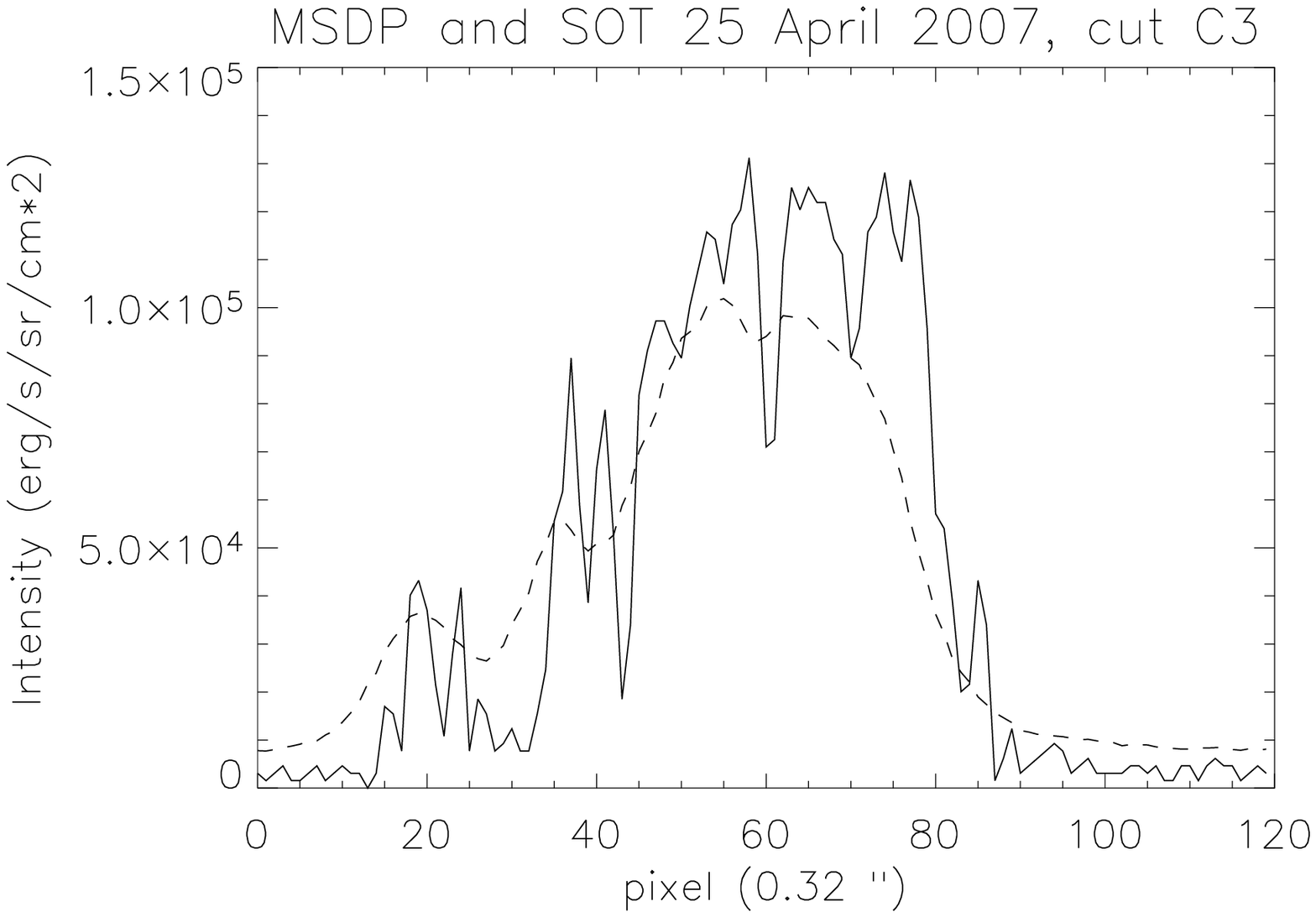}
\caption{Calibrated intensities of cuts (dashed lines) through the
prominence parallel to the limb observed by  MSDP from the South to
the North  overlaid by cuts (solid lines) obtained through the SOT
image of the same prominence at the same time (13:19 UT) for 3
different altitudes 7.5, 18.7, 30 Mm above the solar limb. Cuts 1 is
below the main prominence and crosses the bubbles. It is more
extended than the other cuts. The cut locations are indicated in
Fig. \ref{msdp_sot}.} \label{cuts}
\end{figure}

\subsection{\emph{\emph{Hinode}}/SOT observations}

The \emph{Hinode} mission  is operating since October 2006 \citep{Kosugi07}.
The prominence under study
was well observed by the \emph{Hinode}/SOT instrument between 13:04 and
13:59 UT in H$\alpha$  and in Ca II H lines on April 25, 2007. The
50 cm diameter SOT can obtain a continuous, seeing-free series of
diffraction-limited images in 388-668 nm range with 0.2--0.3
arcsec spatial resolution. The field-of-view of CaII H line is
smaller (108 $\times$ 108 arcsec) and does not cover the whole prominence.
We are using in  our study only H$\alpha$ images (160 $\times$ 160 arc sec)
registered as  a 1024$\times$1024 pixels matrix, each pixel having
dimensions of 0.16 $\times$ 0.16 arcsec.
{\bf The SOT NFI filter is centered on the H$\alpha$ line (as
determined in a line scan calibration prior to the observations)
with a bandpass width of 120 m\AA.} Prominence with Dopplershifts
larger than 20 km s$^{-1}$ can not be observed because the maximum
intensity would be out of the bandpass. The center of the
field-of-view in solar coordinates was [830,-510] arcsec and the
exposure time was 300 ms. The images have been dark-subtracted and
flat-fielded to remove CCD fringes in the H$\alpha$ images (Fig.
\ref{msdp_sot}). The images have been sharpened by an unsharp mask
procedure in order to increase the fine structure contrast. Looking
at the H$\alpha$ SOT  movie, we observe that  the fine structures of
the prominences evolved very rapidly, principally the round-shaped
structures, dark ``bubbles'' at the bottom of the prominence. They
are rising and with their half circle shapes. The material in the
prominence lying above these dark features are very dynamic as these
features are moving up. The material in the bright structures appear
to move down. We notice from time to time brighter threads
surrounding the top of the bubbles with accelerated velocities. The
dynamics of this prominence is comparable to the hedgerow prominence
described by \citet{Berger08}.

\begin{figure}
   \centering
\includegraphics[width=0.50\textwidth,clip=]{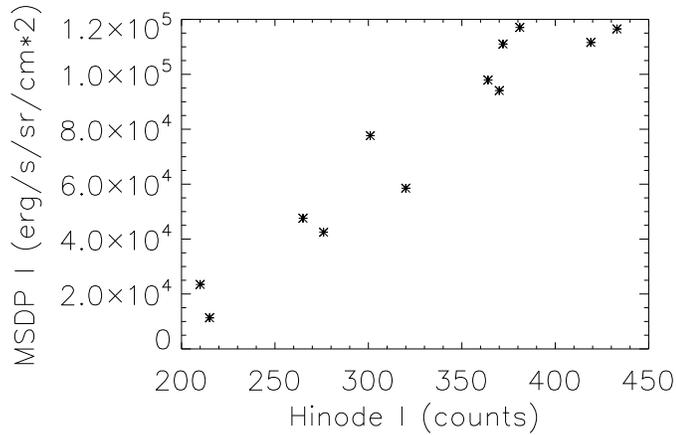}
\caption{Counts of the H$\alpha$ SOT intensity versus calibrated MSDP
integrated intensity in ergs/s/sr/cm$^2$. The plot shows linear behaviour.}
\label{calib}
\end{figure}

\subsection{MSDP observations}

The prominence has been observed during three  consecutive days, on
the April 24 , 25, 26   with the  Multichannel Subtractive Double
Pass spectrograph (MSDP) operating in the solar tower of Meudon. On
April 25 the prominence was observed  between 12:09   and  13:32 UT.
The entrance field stop of the spectrograph covers an elementary
field-of-view of  72$\times$465 arcsec with a pixel size of 0.5 arcsec.
The  final field-of-view of the images is 500$\times$465 arcsec. The
exposure time is 250 ms. We performed consecutive sequences of 60
images with a cadence of 30 sec. The spatial resolution is estimated
to be between of 1 arcsec and 2 arcsec depending on the seeing.
Using the MSDP technique \citep{Mein77,Meinp91}  the H$\alpha $
image of the field-of-view is split in wavelength into nine images
covering the same field of view. The nine images are recorded
simultaneously on a CCD Princeton  camera. Each image is obtained in
a different wavelength interval. The wavelength separation between
the center wavelength of one image to the next is 0.3 \AA.  By
interpolating with spline functions between the observed intensity
in these images, we are able to construct H$\alpha$ profiles in each
point of the observed field-of-view. A mean or reference disk
profile is obtained by averaging over a quiet region on the disk in
the vicinity of the prominence (this case at sin$\theta$ =0.98). The
photometric calibration is done by fitting the reference profile to
standard profiles for the quiet  Sun \citep{David61}. We corrected the
profile of the scattering light by looking at the  nearby corona.
The observations of the prominence in H$\alpha$ line center by the
MSDP spectrograph are easily coaligned with the H$\alpha$ SOT images
obtained at the same times  (Fig. \ref{msdp_sot}).

\begin{figure}
   \centering
\includegraphics[width=0.45\textwidth,clip=]{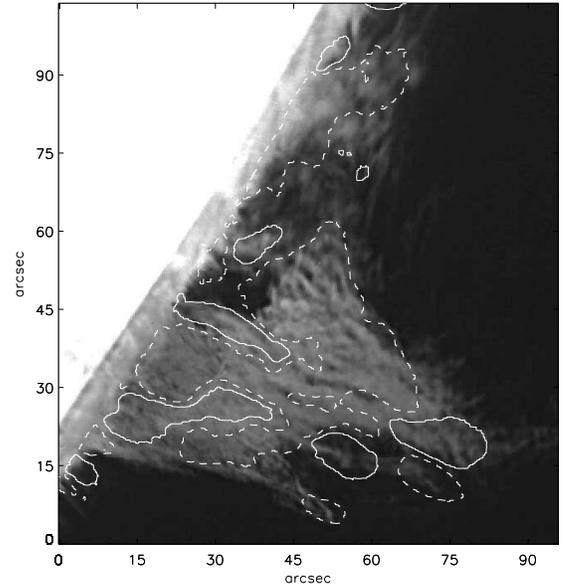}
\includegraphics[width=0.45\textwidth,clip=]{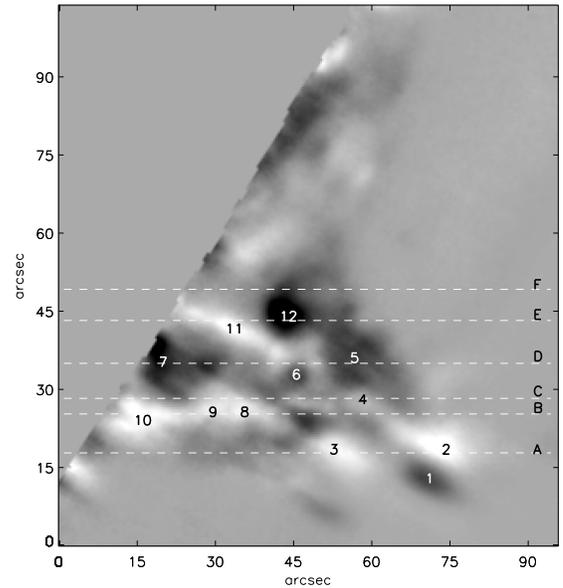}
\caption{Top: H$\alpha$ \emph{Hinode}/SOT image at 13:19 UT overlaid by
Dopplershift contours, bottom: Dopplershift map of MSDP data  at the
same time. White/black areas (solid/dashed lines) correspond
respectively to blue/redshifts. The dashed straight lines in the
bottom image are the location of the slices used to compute the
transverse velocity (A to F). The numbers are the points where
Dopplershifts have been computed (see Table 1). Some profiles have
been drawn in Fig. \ref{prof}.  } \label{doppler}
\end{figure}

\subsection{Normalization of the  H$\alpha$ intensities}

The normalization of H$\alpha$ intensity  allows the  comparison of
observations with  the theory provided by  radiative transfer codes
\citep{Gouttebroze93,Heinzel05} leading to
the determination of physical quantities of prominences. This is
not the scope of this paper. Nevertheless it is interesting to
calibrate the \emph{Hinode} data using the MSDP data.
The intensities of the MSDP observed
profiles are normalized to the local continuum $ I_{\lambda} /
(I_{c,loc})_{obs}$,  the intensities of the local continuum  to the
continuum at the disk center $ I_{(c,loc)}/ (I_c)$. The continuum
at disk center in the wavelength region close to the H$\alpha$ line is {\bf \citep{David61}}:\\
$I_c$ = 4.077 x $10^{-5}$  erg cm$^{-2}$  s$^{-1}$  sr$^{-1}$  Hz$^{-1}$ \\
To perform the  normalization we must
 apply two corrections : (i) one due to the limb darkening, the reference profiles are measured
on the disk near the limb at sin$(\theta)$= 0.98. It reduces the
intensity by a factor 0.547, (ii) the second one taking into account
the $\theta$ angle,  the central intensity is no longer equal to
16$\%$ of the intensity of the continuum but is equal to 22.6 $\%$
of the intensity of the continuum for sin$(\theta)$= 0.98. We
present cuts drawn parallel to the limb at 3 different altitudes
over the limb through the MSDP prominence (see the positions of the
cuts in Fig. \ref{msdp_sot} top panel). The SOT prominence cuts at
the same positions have been over-plotted to the MSDP cuts (the
bubbles have a contrast between 70 to 90 $\%$) (Fig. \ref{cuts}).
The MSDP cuts are smooth with lower contrasts compared to SOT cuts
because of the seeing. This allows us to calibrate H$\alpha$ SOT
observations. This is valid with some possible translation {\bf
according to the accuracy of the wavelength determination}. Using
the count numbers of the SOT Level-1 file, we derive the calibration
curve shown in Fig. \ref{calib}. The integrated H$\alpha$ intensity
of bright threads are around 1.5 $\times 10^{5}$ ergs/s/sr/cm$^2$.

\section{Dopplershifts}

\begin{figure}
\centerline{\hspace*{0.015\textwidth}
             \includegraphics[width=0.25\textwidth,clip=]{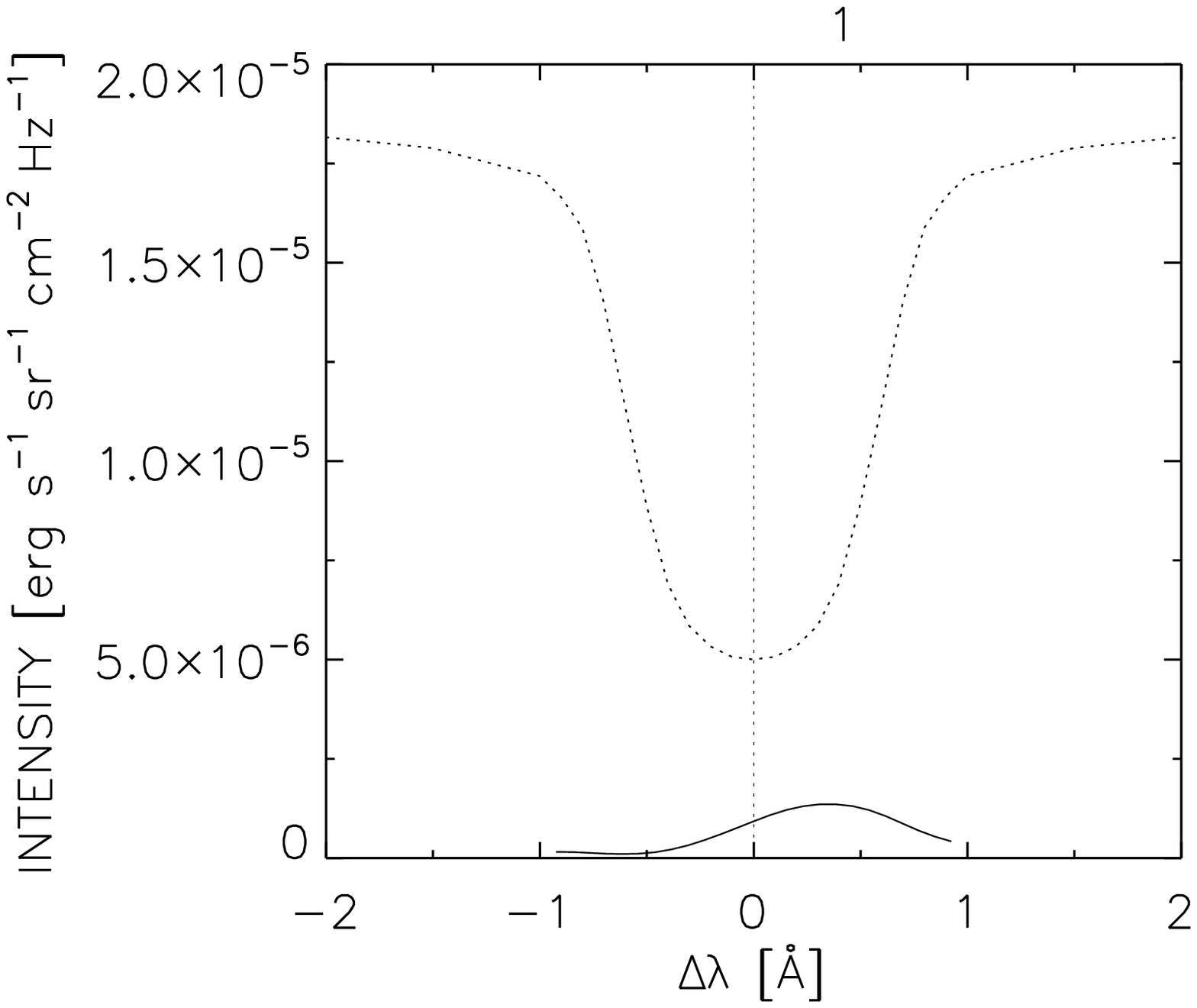}
\includegraphics[width=0.25\textwidth,clip=]{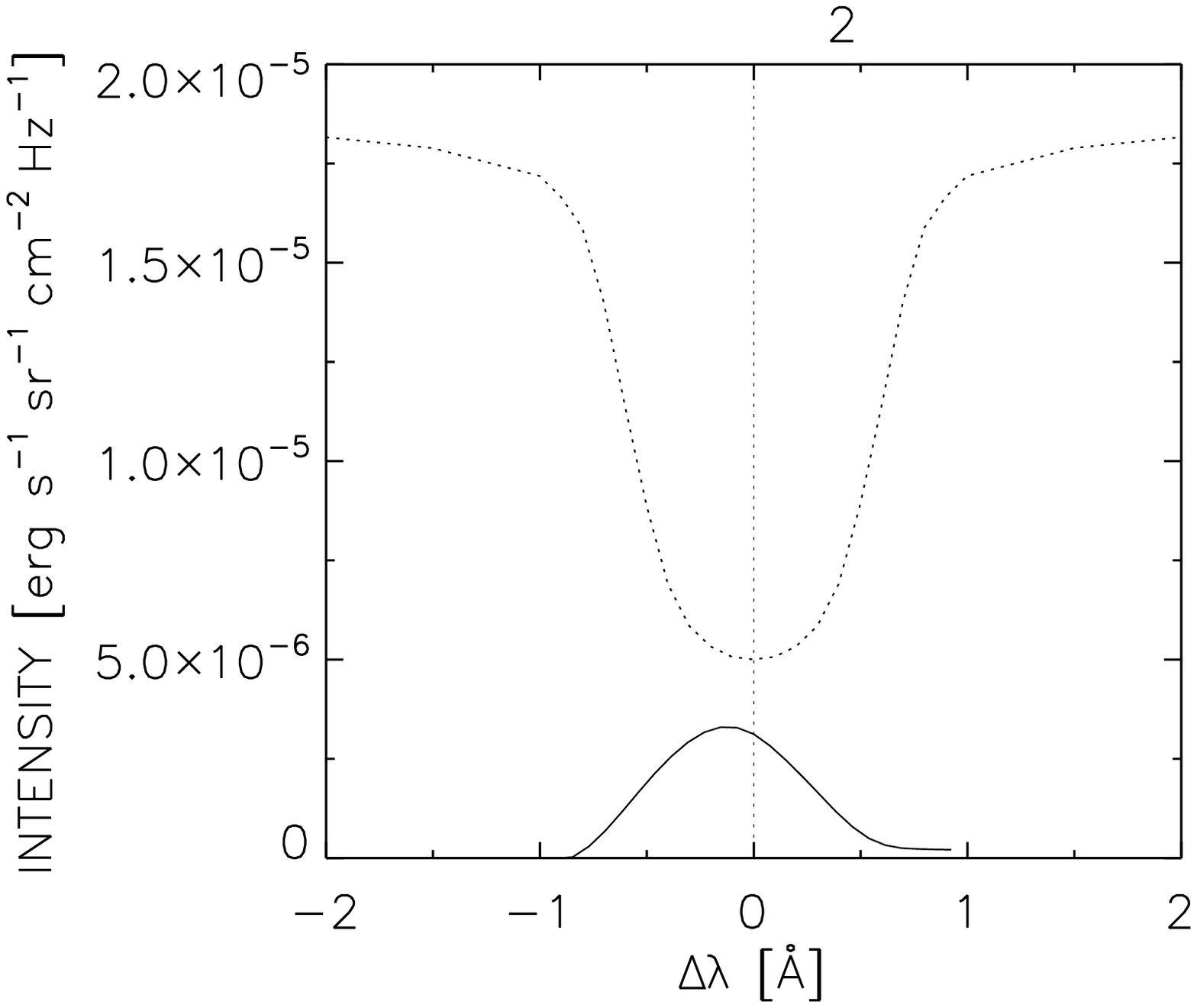}
}
\centerline{\hspace*{0.015\textwidth}
             \includegraphics[width=0.25\textwidth,clip=]{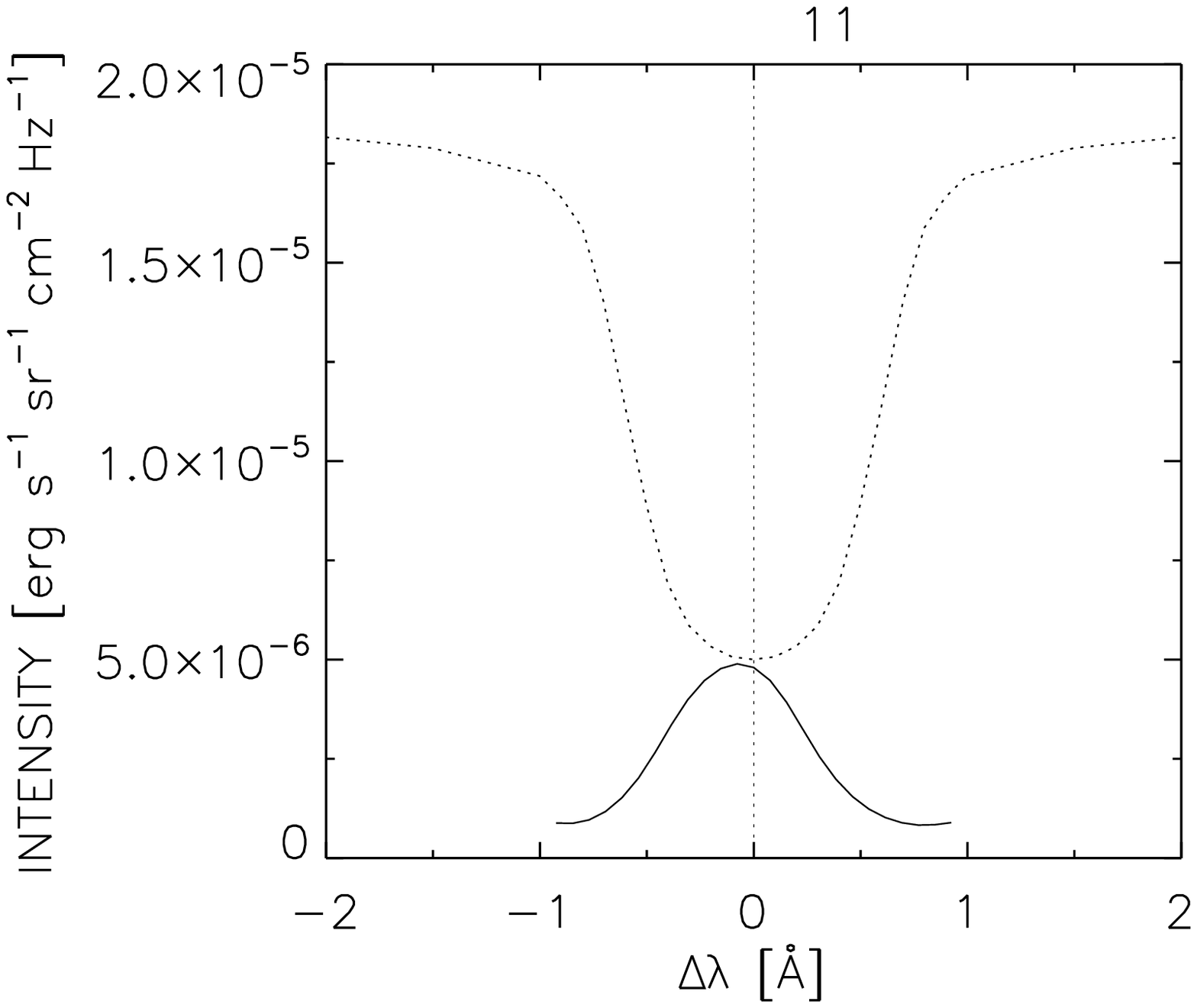}
\includegraphics[width=0.25\textwidth,clip=]{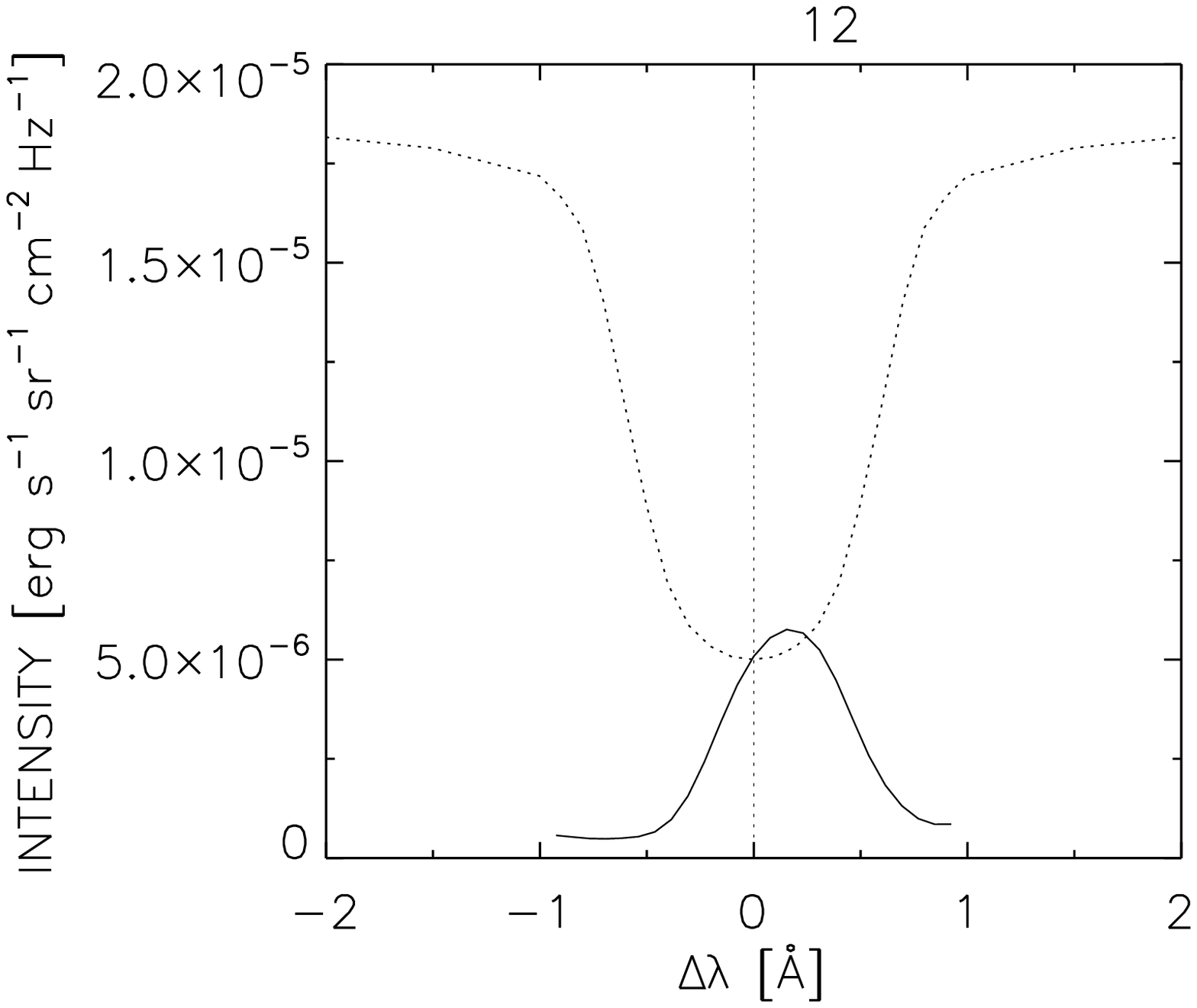}}
\caption{Profiles of the H$\alpha$ line in different points of the
prominence observed with the MSDP at 13:19 UT (in
ergs/s/sr/cm$^2$/Hz). The numbers at the top of panels correspond to
the points marked in Fig. 5. The intensities of profiles of the
prominences (solid lines) have been multiplied by a factor 3 and
compared with the reference line profiles observed on the disk.
The scale in y-axis refers to reference profiles (dotted lines).}
\label{prof}
\end{figure}

A Dopplershift map is  presented in Fig. \ref{doppler}.  We notice
the trend of the velocity pattern with more or less a succession of
vertical blue and redshifts elongated cells/ strands. We compute the
velocity V(y), in a reference system  (x,y,z), (x,z) being the plane
of the sky. We use for that computation the bisector method at the
inflexion point of the line profile ($\pm$ 0.45 \AA). The trend of
this pattern evolves in an hour of observations. Close to the top of
the prominence we measure counterstreaming with redshifts of 15 km
s$^{-1}$ and blue shifts of -5.5 km s$^{-1}$ (Table 1). In the
central part of the prominence the velocity is between $\pm$1 to
$\pm$4 km s$^{-1}$ . At the top of the bubble bright threads move
with a velocity of  7 km s$^{-1}$. The location of the points
mentioned in Table 1 are indicated on the Fig. \ref{doppler}. The
profiles of the H$\alpha$ line are all relatively narrow (Fig.
\ref{prof}). This means that the different threads along the line of
sight have similar velocity {\bf  but they still have somewhat lower
values due to the smearing and seeing effects. } This should also be
true for threads close to each other in the same pixel. It is
difficult to estimate the filling factor. Fine structures are tied
in {\bf bunches}. This remark has been already done for previous
observations (Mein and Mein 1991).

\section{Velocity perpendicular to the line-of-sight}

The \emph{Hinode} movie shows tremendous downflows mainly for the bright
threads and upflows for the dark cavities, `bubbles' rising from the
solar limb. This prominence has a similar behaviour as the
prominence observed in Ca II H by \citet{Berger08}. The time
slice technique \citep{Lin05} allows us  to get quantitative
values of the velocities  V(x,z) or V(trans) in the sky plane  for
different locations in the prominence observed by SOT. We adopt
slices of 5 pixels (equivalent to 0.8$^{\prime\prime}$) with the East-West
orientation (see Fig. \ref{doppler}, bottom). They follow nearly the
common orientation of the fine structures of the prominence. The
maximum angle between the East-West direction  and the fine
structures orientation is  equal to 30 degrees. For these structures
the measured velocities should be multiplied by
 a factor 1.25.  We did not apply systematically this correction because
 for each thread, the angle is different and
 Dopplershifts are also lower values.

\begin{figure}
\vspace*{-1cm}
   \centering
\hspace*{-2cm}
\includegraphics[width=0.60\textwidth,clip=]{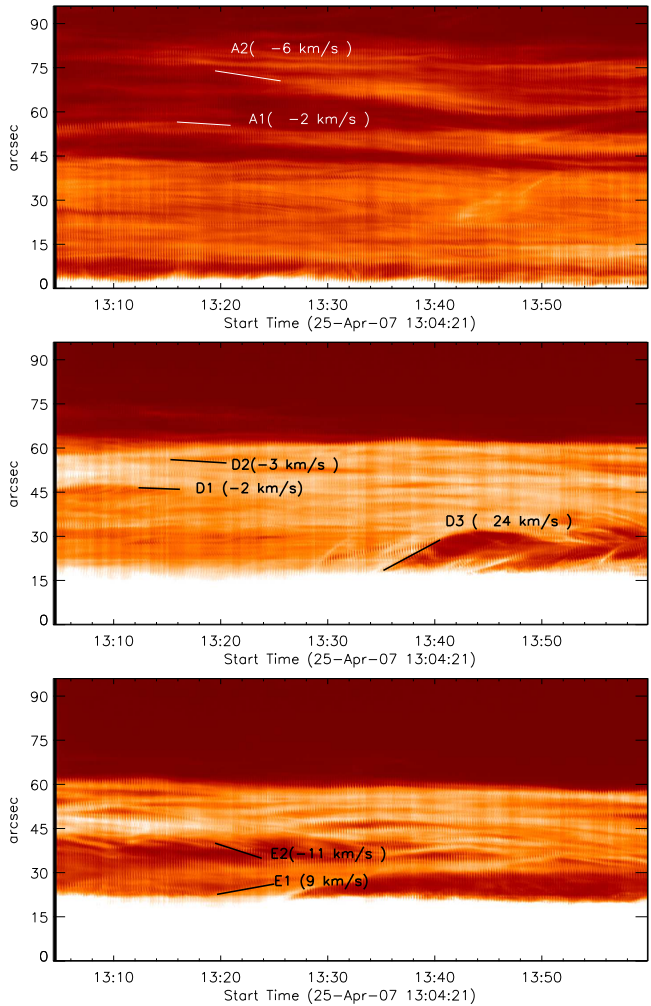}
\vspace*{-1cm}

\caption{Transverse velocities in SOT bright structures using time
slice technique (axis x unit is time, axis y unit is arc sec along
the slide). Top/medium/bottom frame corresponds to slice A/ D/ E,
drawn in Fig. \ref{doppler}. They show the velocities measured
respectively in points A1, A2, D1, D2, E1, E2 (Table 1).
Positive/negative velocities correspond to up/down flows. The large
value +24  km s$^{-1}$ corresponds to the speed of a rising bubble
from the limb or the flow  speed of its bright edge. Fine threads
close to the limb are spicules. Wave  pattern corresponds to
oscillations with 15 to 20 min of period. Adjacent pixels in a slice
have coherent velocities. } \label{sot_velocity}
\end{figure}

 It appears that many structures  in the time-slice diagrams (pixels close to
each other along a slice) have the same velocity. A few of  the threads
show a unique behaviour. Those threads  have in general higher
velocities, reaching -10 km s$^{-1}$ for a short time (5 to 10 minutes).
The other ones have commonly a velocity of the order of -2 to -6
km s$^{-1}$ . The trend is downflows for the bright structures. The
transverse velocity  V(x,z) and the norm of the velocity vector
$\sqrt{[(V(y)^2 +V(x,z)^2)]}$
 are indicated in Table 1. According to the velocity vector, the fine
 structures would be not really vertical but  inclined from the
 vertical by an angle between 30 degrees to 90 degrees.
Some pixels exhibit upflows and later down flows. We identify these
motions as stationary waves with periods of ten to twenty minutes.
If we consider that these pixels belong to  vertical structures, the
plasma is oscillating along/in the structures, if we consider the
structures horizontal, the structures themselves oscillate like the
\citet{Okamoto07} transverse waves. Fig. \ref{sot_velocity}
shows the transverse velocities.

\begin{table}
\caption{Dopplershifts V(Doppler) or V(y) and transverse velocities
V(trans) or V(x,z) in 12 points in the prominence observed with SOT
and the MSDP spectrograph. The 12 points correspond to the named
points in Fig. \ref{doppler} and \ref{sot_velocity}, numbers are
for MSDP data and letters are for SOT data. Positive/negative
Dopplershifts, V(Doppler) are red/blueshifts, positive/negative
transverse velocity, V(trans) are up/down flows. The last column is
the norm of the velocity vector.}\label{table}
\centering
\begin{tabular}{lllll}
&&&&\\
\hline
point & V(Doppler) & V(trans)&  point& $|V|$ \\
  MSDP       & km s$^{-1}$       &km s$^{-1}$ & SOT& km s$^{-1}$  \\
\hline
 1    & +15 & $-$ & $-$ & 15 \\
2& -5.5 & -6& A2& 8.5 \\
3& -6.7 & -2&A1& 7 \\
4 &-0.6 &+4 &C   &  4 \\
5 & +1.8 & -3  &D2& 3.5\\
6&+1.0 &-2 &D1 & 1.4\\
7 & +3.0& $-$ & $-$  & 3.0 \\
8&-1.8 & +3&B3& 3.5 \\
9 & -1.8 &+3&B2& 3.5 \\
10 & -3.7&-2 &B1 &4.2\\
11&-3.6 & +9&E1& 9.7 \\
12 &+7.4 &-11&E2& 13.2\\
\hline

\end{tabular}
\end{table}


\section{Discussion and Conclusion}

For the first time an H$\alpha$ hedgerow prominence has been
observed simultaneously by a high spatial resolution telescope
(\emph{Hinode}/SOT) and by a spectrograph (MSDP) operating in  the Meudon
solar tower on April 25 2007. \emph{Hinode}/SOT allows us to get
the velocities perpendicular to the line-of-sight V(x,z). The second
-- the line-of-sight velocities  V(y) derived from Dopplershifts.
The prominence shows strong dynamics in the SOT movie with dark
cavities rising from the limb with an upward velocity reaching 24
km s$^{-1}$ and downflowing vertical-like bright threads. These threads are
moving horizontally to avoid the dark cavities. During the rise of
cavities, ahead of them,  are observed bright curved fine structures
from time to time with high velocities similar to the speed rise of
the bubble. The \emph{Hinode}/SOT observations have been calibrated by
using the MSDP data. The integrated H$\alpha$ intensity of the
threads reaches 1.5 $\times 10^5$ erg/s/sr/cm$^2$. The contrast of the
dark cavities is between 70 and 90$\%$.

The transverse velocities V(x,z)  of the bright threads are computed
by time slice technique and these values are of the order of  a few
km s$^{-1}$ to 6 km s$^{-1}$ reaching 11 km s$^{-1}$ for individual
fine threads. The pattern of the Dopplershift map show elongated
cells nearly perpendicular to the limb. They are {\bf wider than the
MSDP spatial resolution}. The time slice maps exhibit several pixels
close to each other along the slice having a similar velocity trend.
This means that fine threads close to each other have a coherent
displacement. According to the observations of the prominence three
days before when it is still on the disk as a filament, it appears
that only the feet or barbs are enough dense to be observed. The
prominence would represent the barb threads integrated along the
line-of-sight as the filament is crossing the limb. The structures
are not vertical in the sky plane (x,z) as suggested by the movie.
Dopplershifts and transverse velocities are of same order of
magnitude (less than 6 km s$^{-1}$). In Table 1  we have selected
individual threads with the largest transverse velocities in region
with higher Dopplershifts. The other parts of the prominence exhibit
coherent velocity much smaller (1 to 2 km s$^{-1}$) difficult to
measure. The narrow profiles of H$\alpha$ lines of the prominence
indicate  that the different threads integrated along the line of
sight have similar velocities. The dispersion of the velocities
along the line-of-sight is small.

 The longitudinal magnetic field observed (by the SOHO/MDI instrument) in the filament channel on the disk and on both edges of the inversion line is weak. The strength of the
small polarities are less than 10 Gauss. The prominence lies in a quiet
region and corresponds to a quiescent filament. The small polarities
can change rapidly and this would explain the fast dynamics of the
structures.

In  a flux tube model \citep{Aulanier98,Dudik08} the H$\alpha$
filament  is considered as cool material trapped in shallow dips along
long magnetic  field lines. The feet are extension of the flux tube
disturbed  laterally by parasitic polarities. The barbs are piled up
dips touching the  photosphere. When a parasitic polarity cancels or
moves,  the feet move and can even disappear
\citep{Aulanier98b,Schmieder06,Gosain09}. {\bf In this 3D
perspective
the prominence material would be  trapped in
inclined field lines   and the downflow motion would occur along the
shallow dips.  The brightness would result of the integration of the
threads along the line-of-sight \citep[see Fig. 3e in ][]{Dudik08}.}
\citet{Aulanier98b} explain very well through their magnetic
extrapolation the relationship between parasitic polarities and the
flux tube itself and their evolution. The flux of parasitic polarity
overcomes that of the twisted flux tube and destroys the twisted
configuration. The bubbles would be structures more magnetized that
the surrounding and represented  by the separatrices. A small
increase of  magnetic pressure in the bubble would lead to the rise
of it in the atmosphere. Strong currents can be created in the quasi
separatrice layers (QSL) around the separatrices  by photospheric
displacements of the parasitic polarities \citep{Demoulin96}. Energy
release is expected. This could correspond to the brightening {\bf rims}
associated with filaments \citep{Heinzel95}. They are not
systematically visible due to the dense plasma of filaments. In SOT
observations it could correspond to the bright top edge of the
cavities where reconnection could occur and expel plasma. This would
explain the fast velocity material in brighter threads surrounding
the dark cavities. On the next days, such bubbles are not observed
in the prominence because the feet and parasitic polarities related
to it would be on the back side of the disk.
{\bf In the arcade model with dips proposed by \citet{Heinzel01}
the prominence observed on April 25 may would consist of
vertical threads trapped in dips and piled up giving the impression
of vertical continuous threads. The downflows of 1
to 5 km s$^{-1}$ would be due to shrinkage or successive reconnections of field lines.}

Another explanation for the buoyancy of the dark cavities or bubbles could be
adiabatic expansion of a heated volume of plasma \citep{Berger08}.
This is not exclusive of the magnetic pressure increase scenario and
both magnetic and thermal buoyancy may play a role in the formation
of these structures. We would like to measure the magnetic field in
the bubbles and in the prominence. An interesting aspect would be also
to analyse the EIS and SUMER data to see if the dark bubbles are
filled with hot material. Such measurements are needed to know which
mechanism is valid for the formation of these dark low cavities.


\begin{acknowledgements}
We thank  Nicolas Labrosse  the chief planner of JOP 178 during the
Hinode-SUMER campaign at MEDOC, operating center in Orsay who
forecasted the correct pointing of this prominence three days in
advance. We thank T. E. Berger of the Lockheed Martin Solar and
Astrophysic Laboratory for processing the \emph{Hinode}/SOT data and
for correcting the English language of the manuscript. 
R.C. thanks the CEFIPRA for his post-doctoral grant. This work in done in the frame of the European
network SOLAIRE. We would like to thank the \emph{Hinode} science
team for the observations of SOT, the Meudon solar tower team for
the MSDP observations particularly G. Molodij. We thank Guo Yang for
his help in developing the time-slice procedure. The work of A.B.
was supported by the Polish Ministry of Science and Higher Education
(grant N203 016 32/2287), by the Academy of Sciences of the Czech
Republic (grant M100030942) and the Observatoire de Paris.
 \emph{Hinode} is a Japanese mission developed
and launched by ISAS/JAXA, collaborating with NAOJ as a domestic
partner, NASA and STFC (UK) as international partners.
\end{acknowledgements}

\end{document}